# Cavity-Free Δ-Type Coherent Population Trapping for Microwave Sensing

Ido Fridman, Shemuel Sternklar, and Eliran Talker[*]

Department of Electrical and Electronics Engineering, Ariel University, Ariel 40700, Israel.
*elirant@ariel.ac.il

**Abstract**

We investigated experimentally and theoretically a cavity-free microwave field that couples the two ground states of a Λ-type atomic system, thereby forming a closed Δ configuration. In this regime, the absence of cavity-imposed phase matching leads to a strong sensitivity of the ground-state coherence to the microwave field parameters. We observe that the coherent population trapping (CPT) resonance exhibits a pronounced dependence on the microwave power and detuning, resulting in measurable changes in resonance contrast, linewidth, and center frequency. To explain these effects, we develop a numerical density-matrix model in which the ground-state coherence explicitly incorporates the microwave coupling strength, capturing the essential physics of this no-phase-matching Δ system. The excellent agreement between theory and experiment establishes a simple and robust framework for microwave control of cavity-free Δ-type atomic systems, with direct implications for compact atomic clocks and quantum-enhanced quantum sensing platforms.

**Introduction**

Coherent population trapping (CPT) in multilevel atomic systems is a well-established phenomenon that enables ultra-narrow transparency windows in otherwise absorbing media [1–7]. In standard Λ-type three-level configurations, CPT arises from destructive quantum interference between two optical pathways, and has been extensively exploited in compact atomic clocks [1,3,6–8], slow and stored light [8–11], magnetometry [12–16], quantum memories [4,5 ,11–13], and RF and microwave sensing and imagine schemes [20–26]. More recently, attention has turned to Δ-type systems, in which all three levels are coupled in a closed loop by two optical fields and a microwave (MW) field [20–22,22–24,27,28]. In such configurations, the MW field typically couples the two hyperfine ground states via a weak magnetic-dipole transition. Previous realizations have relied on resonant cavities to enhance this transition and to enforce phase coherence between the optical and MW fields by supporting standing-wave modes inside the cavity [20–22,22–24,27]. In addition, a single RF source is commonly used to drive both the electro-optic modulator (EOM) and the MW antenna, thereby guaranteeing phase locking but at the cost of severely limiting the flexibility and applicability of the system as a practical MW vector sensor [6,8–10]. Such architecture cannot detect uncontrolled external signals and therefore does not faithfully represent realistic sensing conditions. Here we investigate a Δ-type atomic system operating without a resonant cavity and without a common signal generator. The optical fields forming the Λ-system and the external MW field that closes the Δ-loop are generated independently, mimicking realistic scenarios in which the MW signal is not under experimental control and mutual coherence between optical and MW sources cannot be assumed. This cavity-free, free-running configuration allows us to measure fundamental aspects of how CPT responds to an external MW field under these conditions. We show that the additional coupling between the fundamental levels creates a broadening and lowering of the coherent transmission window, which is equivalent to disrupting the coherence between the levels, which we capture in a theoretical model through a MW-power-dependent term. The CPT transparency window is strongly sensitive to the MW power, and that MW frequency detuning relative to the hyperfine transition produces asymmetric shifts of the CPT resonance blue shifts for negative detuning and red shifts for positive detuning. Our finding paves the way for the future use of CPT systems as a basis for vectorial microwave sensors, Similar to various EIT sensing techniques that are capable of measuring an external signal without a resonator, such as the Rydberg atom system that is sensitive to vector components of an MW signal [29–33].

The paper is organized as follows. We first describe the experimental setup and measurement scheme. We then present the main experimental observations of MW-induced modification of the CPT resonance. Next, we develop a numerical model incorporating MW-dependent dephasing in a cavity-free environment and compare its predictions with the data. Finally, we discuss the implications of our results for cavity-free Δ-type systems and for practical microwave vector sensing and conclude.

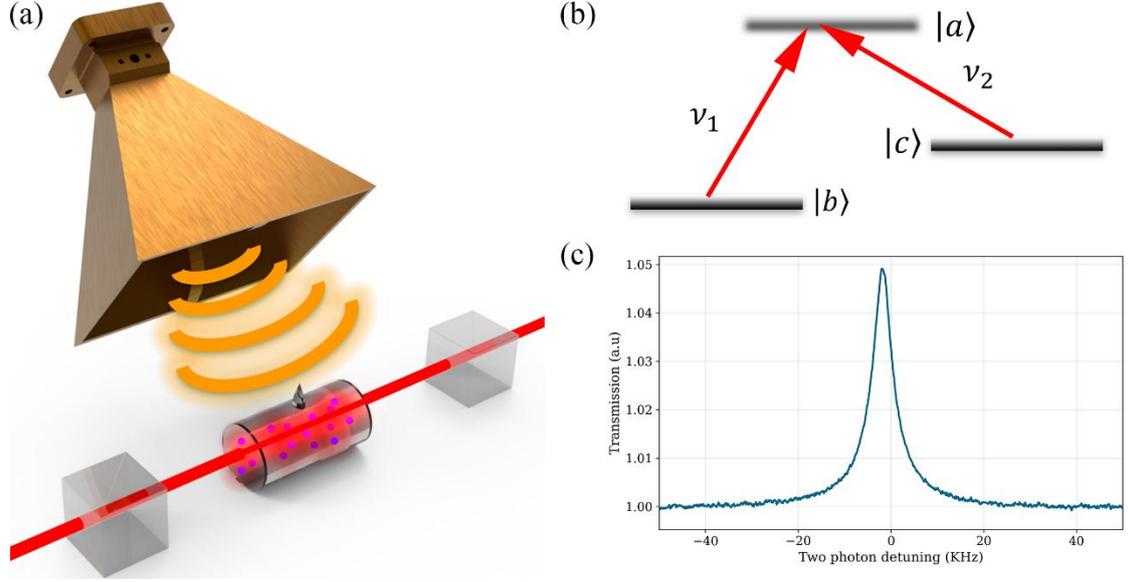

**Figure. 1**. (a) Schematic illustration of the cavity-less experimental configuration. (b) Energy-level diagram of the three-level atomic system. (c) Coherent population trapping (CPT) transmission spectrum obtained from the conventional Λ-type configuration.

**Theoretical model**

Fig. 1(a) illustrates a conceptual image of the Δ-type atomic configuration studied here. In contrast to the conventional Λ-type CPT system shown in Fig.1(b,c), the two ground states are additionally coupled, forming a closed three-level loop. driven by a probe field $E_p$, a coupling field $E_c$, and a cavity-free microwave field $B_\mu$. The optical fields co-propagate along $\hat{z}$,

$$(1) \quad E_j(z,t) = E_j(z)\cos\left(i(\omega_j t - k_j z + \phi_j)\right), j \in \{p, c\}$$

The microwave field propagates along the $x$-axis, while its magnetic component $\mathbf{B}_\mu$ is oriented along the $z$-axis, parallel to the optical quantization axis. This configuration drives $\pi$-polarized Zeeman transitions between the ground-state sublevels. The interaction strength of the microwave field with the atomic system is therefore determined by the projection of the magnetic field along the quantization axis. Accordingly, the microwave magnetic field can be written as

$$(2) \quad B_\mu(x,t) = B_\mu\, e^{i(\omega_\mu t - k_\mu x + \phi_\mu)},$$

where $k_\mu$ is the microwave wavevector accounting for the spatial phase variation of the cavity-free propagating field, $\omega_\mu$ is the microwave angular frequency, and $\phi_\mu$ is the initial phase.

Working in the interaction picture and rotating-wave approximation, with the excited state $|a\rangle$ taken as the reference energy, the Hamiltonian reads

$$(3) \quad H = \Delta_p(v)|b\rangle\langle b| + \Delta_c(v)|c\rangle\langle c| + \Omega_\mu(x,t)|b\rangle\langle c| \\ + \Omega_p(z,t)|b\rangle\langle a| + \Omega_c(z,t)|c\rangle\langle a| + H.c.$$

The Doppler-shifted detunings are

$$(4) \quad \Delta_{p,c}(v) = \omega_{a(b,c)} - \delta_{p,c}(v), \quad \delta_{p,c}(v) = \delta_{p,c} - k_{p,c}v$$

The optical Rabi frequencies are

$$(5) \quad \Omega_p = d_{a,b} \cdot E_p e^{i\phi_p}, \quad \Omega_c = d_{a,c} \cdot E_c e^{i\phi_c}$$

while the microwave Rabi frequency driving the ground-state hyperfine transition is

$$(6) \quad \Omega_\mu(z,t) = \Omega_\pi e^{i((\omega_p - \omega_c - \omega_\mu)t - (k_p - k_c)z + k_\mu x + \phi_\mu)}$$

For the $\pi$-polarized transition $|2,0\rangle \leftrightarrow |3,0\rangle$,

$$(7) \quad \Omega_\pi = \frac{2\mu_B}{\hbar} B_z$$

All Rabi frequencies are taken spatially uniform. To obtain time-independent couplings, we impose the closed-loop resonance condition

$$(8) \quad \omega_p - \omega_c - \omega_\mu = 0$$

And choose $\Delta_p = -\Delta_c$. Unlike $\Lambda$ systems, the propagation phase does not cancel in a $\Delta$ configuration, and the steady state depends explicitly on the relative field phase [21–24].

The density matrix obeys the master equation

$$(9) \quad \dot{\rho}(v,z) = -\frac{i}{\hbar}[H(v,z), \rho(v,z)] + \sum_{k=1}^{5} \mathcal{L}(\hat{c}_k)\rho(v,z)$$

with Lindblad operators

$$(10) \quad \hat{c}_1 = \sqrt{\gamma_{bc}}|b\rangle\langle c|, \hat{c}_2 = \sqrt{\gamma_{cb}}|c\rangle\langle b|, \hat{c}_3 = \sqrt{\gamma_{ab}}|a\rangle\langle b|, \hat{c}_4 = \sqrt{\gamma_{ac}}|a\rangle\langle c|, \\ \hat{c}_5 = \sqrt{\gamma_c}(|b\rangle\langle b| - |c\rangle\langle c|)$$

Here $\gamma_{ab} = \gamma_{ac}$ include spontaneous decay and buffer-gas relaxation of $|a\rangle$, while $\gamma_{bc}$ accounts for ground-state relaxation due to wall collisions, buffer gas, and spin exchange. Since the optical fields are co-propagating, only the longitudinal velocity $v$ contributes to Doppler broadening. The steady-state solution $\rho(v)$ is averaged over the Maxwell–Boltzmann distribution,

(11)
$$\rho = \frac{\int_{-\infty}^{\infty} \rho(v) e^{-\left(\frac{v}{v_{mp}}\right)^2} dv}{\int_{-\infty}^{\infty} e^{-\left(\frac{v}{v_{mp}}\right)^2} dv}$$

Here $v_{mp}$ is the most probable velocity of the atoms and is given by $v_{mp} = \sqrt{3k_B T/m}$ where m is the mass of the rubidium atoms and $k_B$ is the Boltzmann constant. Because the probe and coupling fields originate from the same laser, their initial phases are equal. The relevant closed-loop phase is therefore

(12)
$$\phi(z) = z(k_p - k_c) + k_\mu x + \phi_\mu$$

The response ($\chi$) to the probe signal is proportional to the probe-induced coherence terms $\rho_{ba}$. The relationship is expressed as

(13)
$$\chi = -\frac{N d_{a,b}^2}{\hbar \varepsilon_o \Omega_p} \langle \rho_{ab} \rangle$$

Where N is the density of the Rubidium atoms, $\varepsilon_0$ is the permittivity of free space and $d_{a,b}$ is the moment dipole of transition $a \leftrightarrow b$. For a vapor cell of finite length $L$, the observable response is obtained by accounting for the spatial phase accumulation of $\rho(z)$ across the interaction region.

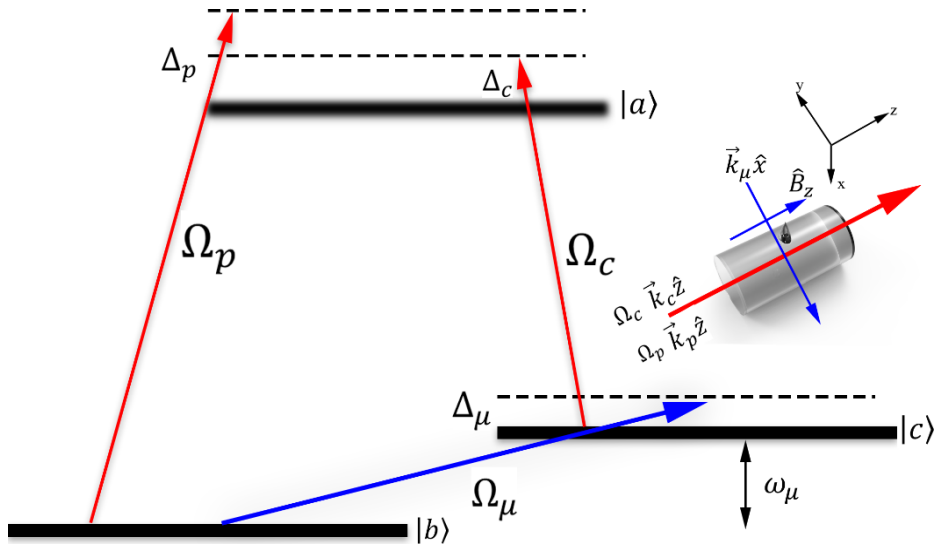

**Figure 2.** Schematic of the Δ-type three-level atomic system. The ground state | b⟩is coupled to the excited state a⟩ by the optical pump field with Rabi frequency $\Omega_p$, while the metastable state | c⟩is coupled to |a⟩ by the optical coupling field with Rabi frequency $\Omega_c$. A microwave field with Rabi frequency $\Omega_\mu$coherently couples the two lower states |b⟩ and |c⟩, completing the closed-loop (Δ-type)

configuration. The corresponding detunings are denoted by $\Delta_p$, $\Delta_c$, and $\Delta_\mu$. The inset illustrates the experimental geometry: the probe and coupling laser beams propagate along the z-axis, while the microwave field propagates along the x-axis, with its magnetic field polarized along z.

To simulate the changes in the probe fields as it passes through a cell of length L, we treat the 85Rb cell as sequence of small cells along the propagation direction. The propagation equation for the probe field is then calculated using the slowly varying envelope approximation in each cell, which is given by [23,24]

$$\frac{\partial \Omega_p}{\partial z} = -i\eta \left( \frac{\omega_\mu N d_{ab}^2}{2\varepsilon_0 c \hbar} \right) \langle \rho_{ba} \rangle \tag{14}$$

Here $\rho_{ba}$ represents the phase-dependent steady-state density matrix element corresponding to probe absorption, in our calculations we set $\eta$ to be close to 1. The phase-dependent matrix elements $\rho_{ba}$ by solving Eqs. ()-() (see Appendix) we can write

$$\rho_{ba} = \frac{i\Gamma_{bc}^0 \Omega_p - \Omega_c \Omega_\mu e^{i(\Delta k \cdot z + k_\mu x + \phi_\mu)}}{\Gamma_{bc}^0 \Gamma_{ba}^0 + |\Omega_c|^2} \tag{15}$$

With $\Gamma_{bc}^0 = \frac{1}{2}\gamma_{bc} + 2\gamma_c - i\Delta_\mu$, $\Gamma_{ba}^0 = \frac{1}{2}\gamma_c + \gamma_{ba} - i\Delta_p$, where

$$\Delta k = k_p - k_c \tag{16}$$

Using the expression for $\rho_{ba}$, we apply the slowly varying envelope approximation for the probe field entering the vapor cell of length L at $z_0$ and exiting at $z_0 + L$. Denoting the initial intensity of the probe field as $\Omega_p^0$, we derive an expression for the intensity of the probe at the exit of the cell to be

$$\Omega_p(x, z_0 + L) = e^{-\alpha L} \left( \Omega_{p0} - i \frac{\alpha \Omega_c \Omega_\mu e^{i\Delta k z_0}}{\Gamma_{bc}^0} \frac{e^{(\alpha + i\Delta k)L + k_\mu x + \phi_\mu} - 1}{\alpha + i\Delta k} \right) \tag{17}$$

With

$$\alpha = \frac{\eta \Gamma_{bc}^0}{\Gamma_{bc}^0 \Gamma_{ba}^0 + |\Omega_c|^2} \tag{18}$$

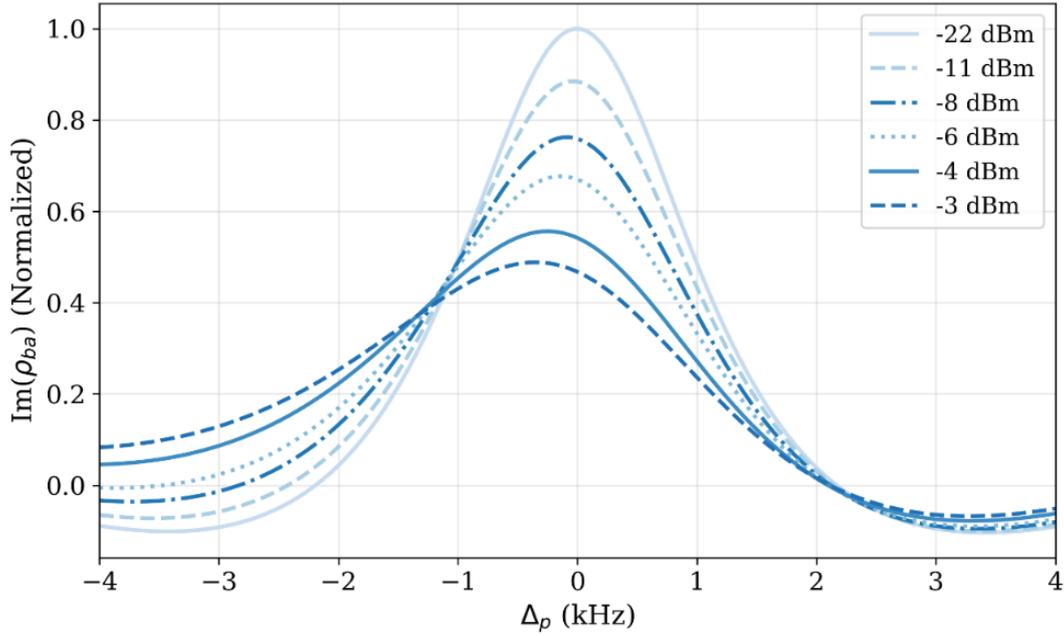

**Figure 3.** Theoretical dependence of the imaginary part of the coherence $\rho_{ba}$, proportional to the probe absorption, as a function of the probe bean detuning at different microwave field power. The results use a three-level dephasing model. In the numerical calculations, the probe and coupling Rabi frequencies are set to $\Omega_p = \Omega_c = 2\pi \times 20$ MHz, and the microwave detuning is fixed at $\Delta_\mu = 0$. $\Gamma_c = 2\pi \times 6.06$ MHz, $\gamma_{bc} = 2\pi \times 15$ kHz

Fig. 3 presents the theoretical dependence of the imaginary part of the coherence $\rho_{ba}$, which is directly proportional to the probe-beam transmission, as a function of the probe detuning. The calculations are performed at zero microwave detuning ($\Delta_\mu = 0$) with a fixed distance of 1 m between the microwave antenna and the rubidium vapor cell.

As the microwave power increases, the absorption spectrum exhibits two systematic effects: a progressive reduction in amplitude and a shift of the resonance position. The decrease in absorption contrast arises from the enhanced microwave-induced coupling between the two ground states, which modifies the steady-state ground-state coherence and redistributes population among the levels. Simultaneously, the observed spectral shift reflects a microwave-induced AC Stark (light-shift–like) effect, whose magnitude increases with the applied microwave field strength. These results highlight the strong sensitivity of the probe absorption to the microwave field amplitude and demonstrate how microwave power directly controls both the strength and spectral position of the CPT-related resonance in the cavity-free Δ-type configuration.

**Experimental Setup**

The experiment is performed using a cylindrical vapor cell containing isotopically enriched [85]Rb, with a length of 30 mm and a radius of 10 mm. The cell is filled with 30

Torr of $N_2$, which serves both as a buffer gas and as a quenching gas. The cell is heated to approximately 57 °C using twisted resistive heating wires driven by an AC current, a configuration that minimizes stray magnetic fields generated by the heating system. The vapor cell is mounted at the center of a three-axis Helmholtz coil assembly, which provides full control over the magnetic field environment. A static magnetic field of approximately 2 G is applied along the optical axis ($\hat{z}$) to define a well-defined quantization axis for the atomic states. This field lifts the Zeeman degeneracy, sufficiently separating the magnetic sublevels so that the system can be accurately treated as an effective three-level system. Additional weak compensating fields are applied along the transverse directions to cancel residual components of the Earth's magnetic field. A schematic illustration of the experimental setup is shown in Fig. 4.

The optical field is derived from a diode laser, whose frequency is stabilized via saturated absorption spectroscopy to the $F = 2 \rightarrow F' = 3$ transition of the $^{85}$Rb D1 transition line. A fraction of the laser output is used for frequency locking, while the main beam is coupled into a single-mode optical fiber and directed to an electro-optic modulator (EOM). The EOM is DC-biased at 5.8 V to operate in the separated-carrier double-sideband (SCDS) regime, generating two coherent optical sidebands asymmetrically detuned from the carrier. The sideband separation is set by a microwave signal generator operating at 1.5178 GHz. To enable controlled scanning of the two-photon detuning, a low-frequency modulation (10 Hz, 18 mV amplitude) is applied to the EOM drive signal, producing a slow frequency scan of several tens of kHz. This results in effective sideband separations ranging from 1.5176 GHz to 1.518 GHz, allowing precise tuning of the ground-state detuning. This scanning scheme enables direct mapping of the optical transmission as a probe function and coupling beam frequency detuning with coherent population trapping (CPT) appearing as a narrow transmission peak near the resonance transition (see Fig. 1c). An acousto-optic modulator (AOM), driven at 80 MHz, is placed after the EOM and is used for active stabilization of the optical power incident on the vapor cell. Following the AOM, the beam is circularly polarized using a quarter-wave plate and directed through the vapor cell. The transmitted light is detected by a photodiode. The system is configured such that the two optical fields—referred to as the pumping ($\Omega_p$) and coupling ($\Omega_c$) beams, are scanned simultaneously with equal magnitude and opposite sign detuning's. At the position of the vapor cell, the beam diameter is approximately 2 mm, and the optical power of each beam is on the order of a few microwatts (150 μW).

Microwave coupling between the ground states is provided by a horn antenna positioned approximately 1 m from the vapor cell. The antenna is oriented such that the magnetic component of the microwave field is aligned parallel to the optical axis. Maximum modification of the CPT transparency window is observed when the microwave frequency is tuned to 3.03574 GHz, corresponding to the ground-state hyperfine splitting of 85Rb. Measurements are performed over a microwave power range from −22 dBm to 2 dBm. The effective microwave power experienced by the atoms is significantly lower than the nominal antenna output power, primarily because only atoms within the small laser-defined interaction volume contribute to the signal and because the microwave field amplitude at the vapor cell is reduced by the antenna radiation pattern and propagation geometry.

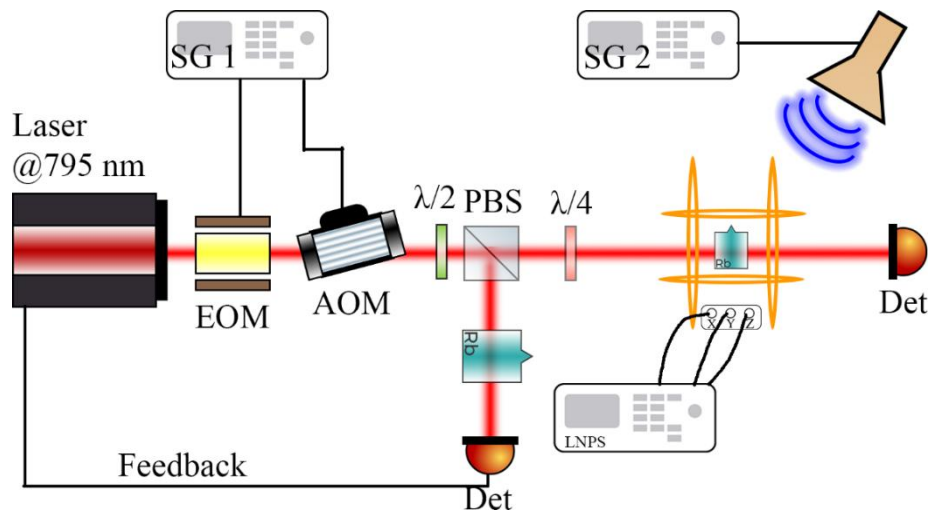

**Figure. 4**. Schematic of the experimental setup is used to perform CPT spectroscopy in alkali vapor cell. DFB - distributed feedback diode laser at wavelength of 795 nm, EOM - Mach–Zehnder electro-optic modulator; AOM - acousto-optic modulator, $\lambda/2$ - half wave plate, $\lambda/4$ - quarter wave plate, LNPS - low noise power supply, PBS - polarizing beam splitter, SG - signal generator, Det - detector.

**Results**

In a Λ-type system, under appropriate conditions, a narrow CPT transmission window is formed, with a full width at half maximum (FWHM) of about 1.5 kHz. When an external microwave field is applied at a frequency close to hyperfine ground states frequencies of $|F = 2, m = 0\rangle \rightarrow |F' = 3, m' = |0\rangle$ the coherence between the levels is disturbed and electron population change is created, resulting in changes on the CPT resonance. In particular, the maximum transmission decreases, and the shape of the transmission window expands, Similar to the change caused by decoherence of the levels. In the experimental setup, an external horn antenna directs a traveling electromagnetic wave perpendicular to the optical axis ($\hat{x}$), with the magnetic component polarized parallel to the optical axis ($\hat{z}$).

Fig. 5 Measured transmission map of the coherent population trapping (CPT) resonance in an $^{85}$Rb vapor cell as a function of the two-photon detuning and the applied microwave power $P_\mu$. As microwave power increases, the coupling between the two ground states progressively degrades the CPT coherence, leading to a systematic reduction in the height (contrast) of the transmission window. This behavior reflects microwave-induced decoherence of the ground-state superposition and the associated suppression of the CPT dark state. The transmission amplitude is normalized to the maximum transmission obtained in the case of $\Delta_\mu = 0$ without the microwave field, which corresponds to approximately 34% of the incident laser power. A reduction in transmission is demonstrated as the power increases.

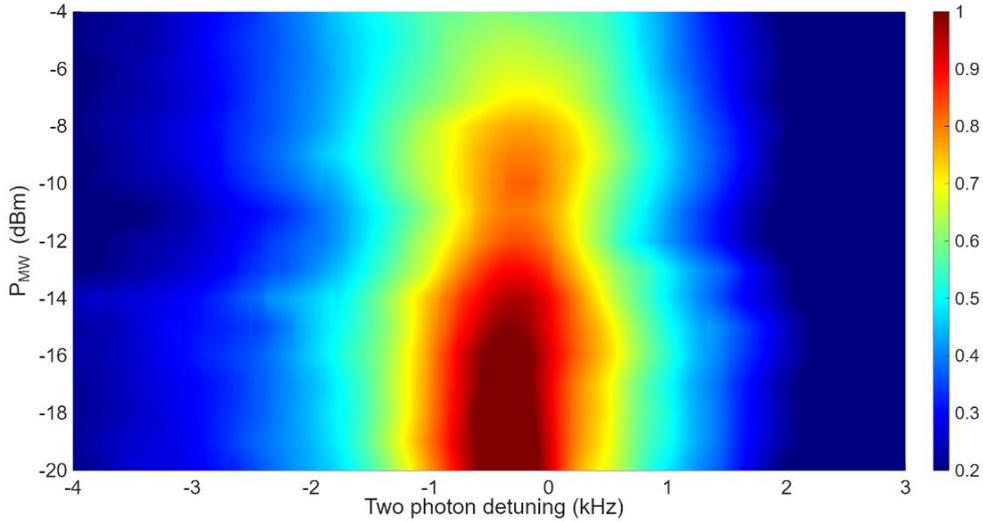

**Figure 5**. CPT-induced transparency window in $^{85}$Rb vapor cell as a function of two-photon detuning and microwave power $P_\mu$. Increasing microwave power reduces the CPT coherence, resulting in a decreased transmission contrast.

Next, we investigated the response of the CPT resonance to detuning of the microwave field relative to the ground-state hyperfine transition frequency. Fig. 6 shows the CPT transmission as a function of the microwave detuning $\Delta_\mu$ at fixed microwave power. A positive microwave detuning ($\Delta_\mu > 0$) preferentially enhances the red-shifted CPT peak, while a negative detuning ($\Delta_\mu < 0$) enhances the blue-shifted peak. This sign-dependent asymmetry originates from a microwave-induced AC Stark shift of the ground-state levels, which modifies the CPT dark-state condition in a frequency-dependent manner. As a result, both the magnitude and sign of the microwave detuning are directly encoded in the CPT line shape, providing a dispersive-like response suitable for frequency discrimination. Notably, the transition from a symmetric to an asymmetric CPT profile occurs over a narrow detuning range of approximately 100 Hz, demonstrating the high intrinsic frequency sensitivity of the system to small deviations of the microwave field from resonance. This sharp response establishes a clear pathway for using the CPT resonance as a microwave frequency discriminator or sensor, without the need for resonant cavities or phase-locked microwave–optical architectures.

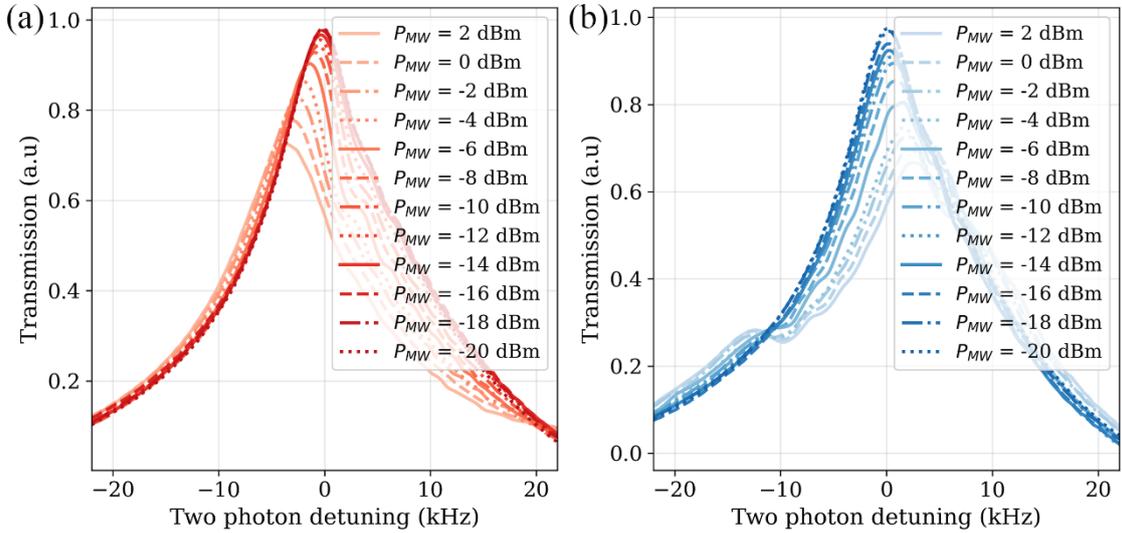

**Figure 6.** CPT spectral response under interaction with a detuned microwave field near the ground-state hyperfine transition. (a) For positive microwave detuning ($\Delta_\mu > 0$), the CPT transmission peak is shifted toward lower optical frequencies (red shift). (b) For negative microwave detuning ($\Delta_\mu < 0$), the CPT peak shifts toward higher optical frequencies (blue shift). This asymmetric frequency shift arises from a microwave-induced AC Stark (light-shift) effect, which modifies the ground-state coherence and shifts the interference condition for CPT transparency. The sign-dependent response directly encodes the magnitude and direction of the microwave detuning, highlighting the potential of the system for precision microwave frequency sensing.

A similar frequency-dependent behavior is observed for several values of $\Delta_\mu$ when the magnetic component of the microwave field is aligned parallel to the optical axis ($\hat{z}$), as shown in Fig. 7, indicating that the sensing mechanism is robust against changes in microwave field orientation. Together, these results demonstrate the feasibility of a compact, cavity-free CPT-based platform for precision microwave frequency sensing and metrological applications.

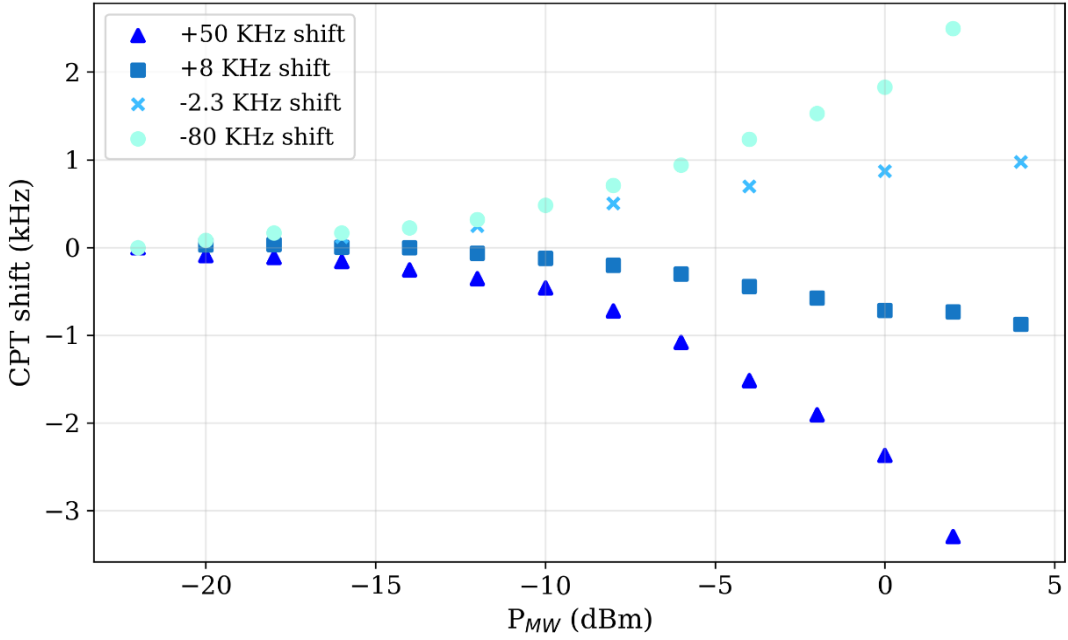

**Figure 7.** Spectral shift of the CPT transmission peak as a function of applied microwave power, measured for four different values of the microwave detuning $\Delta_\mu$. In this configuration, the magnetic component of the microwave field is aligned parallel to the optical axis ($\hat{z}$). The power-dependent shift reflects a microwave-induced AC Stark effect on the ground-state coherence, with the magnitude and direction of the shift determined by $\Delta_\mu$, demonstrating controllable and detuning-sensitive microwave–optical coupling.

In Fig. 8, we compare the experimental measurements with the theoretical model described by Eq. (15). In the model, the Rabi frequencies of the probe and coupling fields are set equal, $\Omega_p = \Omega_c = 2\pi \times 20\,\text{MHz}$, and the microwave detuning is fixed at $\Delta_\mu = 0$. Under these conditions, we observe very good agreement between the numerical simulations and the experimental data over the full range of applied microwave powers. By varying the microwave field intensity, both the experiment and the model show a systematic reduction in the contrast of the CPT transmission feature, as illustrated in Fig. 8. This behavior arises from microwave-induced modification of the ground-state coherence and is analogous to a power-broadening–like effect, in which increasing field strength leads to reduced transmission contrast and partial degradation of the CPT dark state.

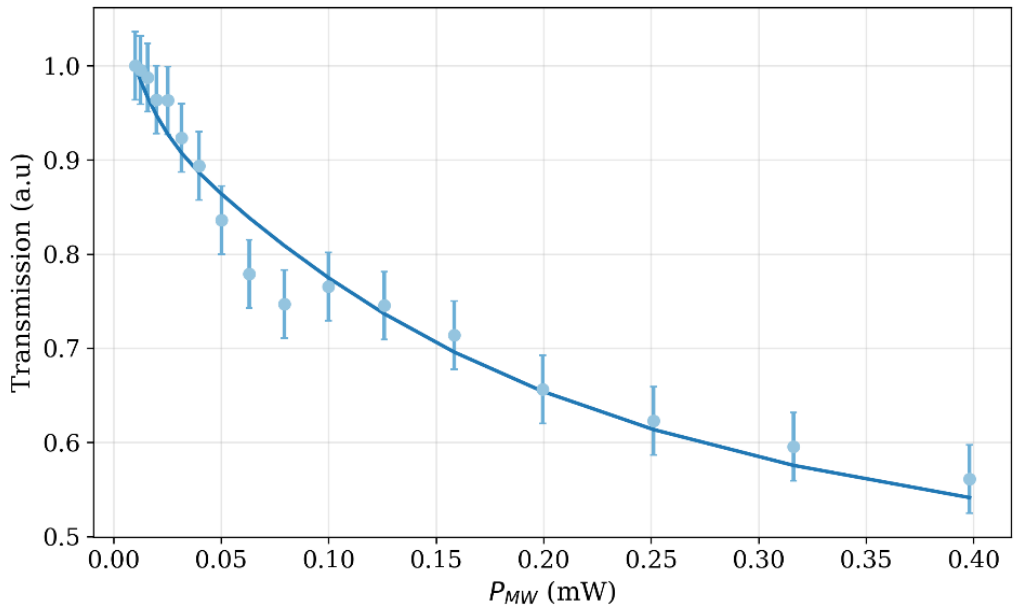

**Figure 8.** CPT transmission contrasts versus microwave field intensity. The measured reduction in contrast with increasing microwave power is well reproduced by the theoretical model, demonstrating quantitative agreement between experiment and theory.

Fig. 9 illustrates the shift of the CPT signal as a function of the microwave intensity for different microwave shift from resonance where in Fig. 9(a) we shift the frequency of the microwave to red shift, and in Fig. 9(b) we shift the frequency to blue shift. We can see according to equation (15-17) that the shift for the case of blue shift detuning is higher than the shift for the case of red shift.

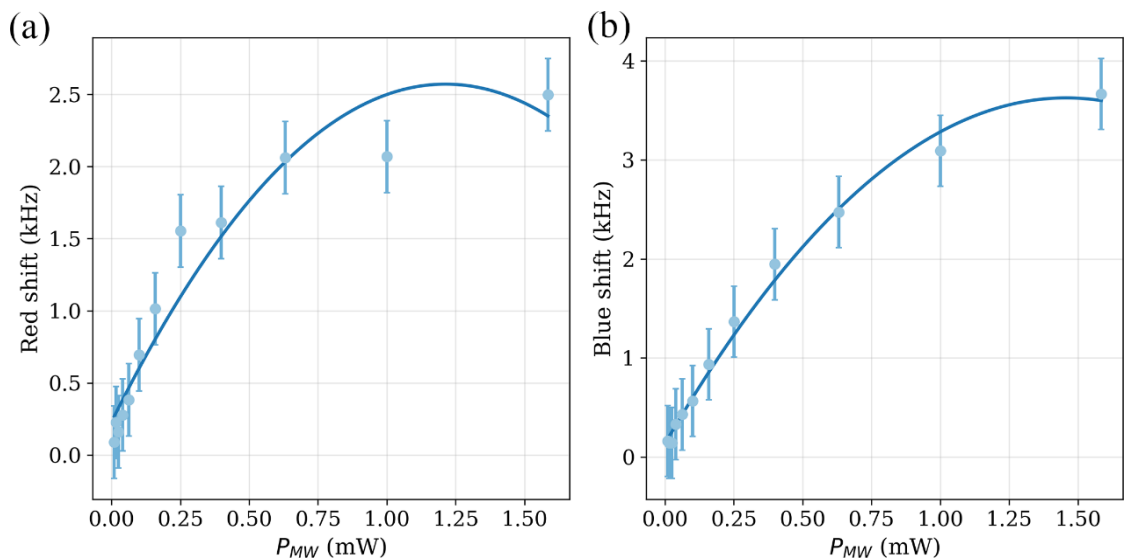

**Figure 9.** Transmission shift and corresponding fit as a function of microwave field intensity for different microwave detuning's: (a) red-detuned microwave excitation and (b) blue-detuned microwave excitation. The data illustrates the distinct dependence of the CPT resonance shift on the microwave power for opposite detuning signs.

## Conclusion

In this work, we presented a combined theoretical and experimental investigation of a three-level Δ-type atomic system driven by an external microwave field under realistic, cavity-free conditions. We demonstrated that the coherent population trapping (CPT) response is strongly influenced by both the amplitude and frequency of the applied microwave field. Owing to the relatively strong static magnetic field ($\approx 2\ G$), the atomic system can be accurately described by an effective three-level model, in which the influence of the microwave field is captured through a phenomenological dephasing term dependent on the microwave Rabi frequency. This model successfully reproduces the experimentally observed behavior in the regime of near-zero microwave detuning ($\Delta_\mu \approx 0$), particularly when the microwave magnetic field component is aligned with the optical axis.

A key advantage of the presented approach is the elimination of any metallic resonator or cavity, which typically isolates the sensor from its electromagnetic environment. In addition, our system is dependent on a common signal generator whose function is to ensure coherence of all signals and maintain a uniform phase, which makes the experimental system theoretical and useless in measuring signals that are not controlled by the experimenter. By operating in an open configuration, our system directly probes the spatial properties of the applied microwave field, enabling sensitivity to external signals without perturbing their spatial distribution. This feature represents an important step toward the realization of compact, open-space vector microwave sensors. While the present work demonstrates sensitivity to a single vector component of the field, it lays the foundation for future studies targeting full vector reconstruction, including direction of arrival, polarization ellipticity, and spatial field gradients. This technique can take advantage of the advantages of millimeter rubidium cells, which have an extremely narrow bandwidth and high sensitivity **[34–36]**.

Finally, we observe a nontrivial dependence of the CPT resonance on the microwave frequency. In addition to the expected AC Stark (light) shift—manifested as a detuning-dependent displacement of the CPT resonance—we find that the magnitude of this shift exhibits a complex dependence on the microwave parameters. This behavior cannot be fully captured by the simplified three-level model employed here, indicating the need for more comprehensive theoretical treatment. These results highlight both the richness of the underlying physics and the potential of cavity-free Δ-type systems for advanced microwave sensing and quantum-enabled field measurements.

**Data availability.** Data underlying the results presented in this paper are not publicly available at this time but may be obtained from the authors upon reasonable request.

# Appendix A : Lindblad master equation

The energy-level structure of the closed Δ-type system is illustrated in Figure 2. The excited state $|a\rangle$ corresponds to the $5P_{3/2}, F' = 3$ level of the $^{85}$Rb atom, while the two ground states $|b\rangle$ and $|c\rangle$ correspond to the hyperfine levels $5S_{1/2}, F = 2$ and $5S_{1/2}, F = 3$, respectively. The energies of these states are denoted by $\hbar\omega_i (i = a, b, c)$.

The optical probe and coupling fields drive the electric-dipole–allowed transitions $|b\rangle \to |a\rangle$ and $|c\rangle \to |a\rangle$, respectively. The ground-state transition $|b\rangle \leftrightarrow |c\rangle$, which is electric-dipole forbidden, is coherently driven by a microwave (MW) magnetic field at 3.035 GHz. The atoms are assumed to move randomly in all directions at room temperature.

Both the probe and coupling laser beams co-propagate along the $\hat{z}$ direction, while the microwave field is applied along the $\hat{y}$ direction. In this closed Λ system, the probe and coupling optical fields are denoted by $\varepsilon_p$ and $\varepsilon_c$, respectively, and the microwave field is denoted by $\varepsilon_\mu$. These fields can be expressed as:

(A1) $$\varepsilon_p = E_p \cos(\omega_p t - k_p z + \phi_p)$$

(A2) $$\varepsilon_c = E_c \cos(\omega_c t - k_c z + \phi_c)$$

(A3) $$\varepsilon_\mu = E_p \cos(\omega_v t - k_\mu x + \phi_\mu)$$

The probe and coupling optical fields have angular frequencies $\omega_p$ and $\omega_c$, and corresponding wave vectors $\mathbf{k}_p$ and $\mathbf{k}_c$, respectively. The microwave field is characterized by its angular frequency $\omega_\mu$, wave vector $\mathbf{k}_\mu$, and phase $\phi_\mu$.

The dynamics of the closed Δ-type system are described using the density matrix formalism. The time evolution of the density operator $\rho$ is governed by the master equation

(A4) $$\dot{\rho} = -\frac{i}{\hbar}[H_{tot}, \rho] + \sum_{j=c,b} \gamma_{aj} \mathcal{L}_{\rho_{ja}} \rho + \frac{\gamma_d}{2} \mathcal{L}_{\rho_d} \rho$$

Here, $\rho$ represent the density operator. The first term in equation () corresponds to the commutator of the total Hamiltonian with the density operator $\rho$. The second term accounts for the damping due to natural relaxation rates $\gamma_{ja}$ (where $j = b, c$) of the electrons from the excited state to the ground states $|b\rangle$ and $|c\rangle$. The

Lindblad operator expansion is given by $\mathcal{L}_{\rho_{ja}}\rho = \frac{1}{2}(2\rho_{ja}\rho\rho_{ja}^{\dagger} - \rho_{ja}^{\dagger}\rho_{ja}\rho - \rho\rho_{ja}^{\dagger}\rho_{ja})$ with the $j = b, c$. The third term represents the phase damping between the two lower levels, with the decoherence rate $\gamma_d$ and the corresponding Lindblad operator is $\mathcal{L}_{\rho_d}\rho = \frac{1}{2}(2\rho_d\rho\rho_d^{\dagger} - \rho_d^{\dagger}\rho_d\rho - \rho\rho_d^{\dagger}\rho_d)$ with the $\rho_d = (\rho_{cc} - \rho_{bb})$. The equations of motion for the population and coherence terms are,

(A4) $\quad \dot{\rho}_{aa} = -(\gamma_{ac} + \gamma_{ab})\rho_{aa} + i\Omega_c(\rho_{ac} - \rho_{ca}) + i\Omega_p(\rho_{ab} - \rho_{ba})$

(A5) $\quad \dot{\rho}_{bb} = -(\gamma_{ab} + \gamma_{ac})\rho_{aa} + i\Omega_c(\rho_{ac} - \rho_{ca}) + i\Omega_p(\rho_{ab} - \rho_{ba})$

(A6) $\quad \dot{\rho}_{bb} = -\gamma_{ab}\rho_{aa} + i\Omega_\mu(e^{i\phi_r}\rho_{bc} - e^{-i\phi_r}\rho_{cb}) + i\Omega_p(\rho_{ba} - \rho_{ab})$

(A7) $\quad \dot{\rho}_{cc} = -\gamma_{ca}\rho_{aa} + i\Omega_\mu(e^{-i\phi_r}\rho_{cb} - e^{i\phi_r}\rho_{bc}) + i\Omega_c(\rho_{ca} - \rho_{ac})$

(A8) $\quad \dot{\rho}_{ab} = -\Gamma_{ab}\rho_{ab} + i\Omega_p(\rho_{aa} - \rho_{bb}) - i\Omega_c\rho_{cb} + i\Omega_\mu e^{i\phi_r}\rho_{ac}$

(A9) $\quad \dot{\rho}_{ac} = -\Gamma_{ac}\rho_{ac} + i\Omega_p(\rho_{aa} - \rho_{bb}) - i\Omega_p\rho_{bc} + i\Omega_\mu e^{-i\phi_r}\rho_{ab}$

(A10) $\quad \dot{\rho}_{cb} = -\Gamma_{cb}\rho_{cb} + i\Omega_\mu e^{i\phi_r}(\rho_{cc} - \rho_{bb}) - i\Omega_c\rho_{ab} + i\Omega_p\rho_{ca}$

The complex decay rates appearing in the optical Bloch equations are defined as

(A11) $\quad \Gamma_{ab} = \frac{1}{4}[2(\gamma_{ab} + \gamma_{ac}) + \gamma_d] + i\delta_p, \Gamma_{ac} = \frac{1}{4}[2(\gamma_{ab} + \gamma_{ac}) + \gamma_d] + i\delta_c,$

And

(A12) $\quad \Gamma_{cb} = \gamma_d + i\delta_\mu,$

where $\delta_p$ and $\delta_c$ denote the detunings of the probe and coupling fields, respectively, and $\delta_\mu$ is the microwave detuning.

The diagonal density-matrix elements $\rho_{aa}$, $\rho_{bb}$, and $\rho_{cc}$ represent the populations of the excited state $|a\rangle$ and the ground states $|b\rangle$ and $|c\rangle$, respectively, while the off-diagonal elements describe the corresponding atomic coherences. The population dynamics are governed by the time derivatives $\dot{\rho}_{aa}$, $\dot{\rho}_{bb}$, and $\dot{\rho}_{cc}$, whereas the coherence dynamics are described by $\dot{\rho}_{ab}$, $\dot{\rho}_{ac}$, and $\dot{\rho}_{cb}$.

The relative phase accumulated by the three interacting fields is given by

(A13) $\quad \phi_r = (k_p - k_c)z - k_\mu x - \phi_\mu,$

where $k_p$ and $k_c$ are the wave vectors of the probe and coupling fields propagating along the z-axis, $k_\mu$ is the microwave wave vector perpendicular to the optical-axis, and $\phi_\mu$ is the microwave phase.

The probe and coupling detunings are defined as $\delta_p = \omega_p - \omega_0$ and $\delta_c = \omega_c - \omega_0$, respectively. At room temperature, the atomic ensemble possesses a nonzero thermal velocity distribution, leading to Doppler broadening of the optical transitions. This effect is incorporated by averaging the optical coherences over the Maxwell–Boltzmann velocity distribution, yielding the velocity-averaged density-matrix elements.

(A14)
$$\langle \rho_{ab} \rangle = \frac{1}{\sqrt{\pi v_d^2}} \int_{-\infty}^{\infty} \rho_{ab}(v) e^{-\frac{v^2}{v_d^2}} dv$$

The average value is calculated by replacing the detuning parameters as $\delta_p(v) = \delta_p - k_p v$, $\delta(c) = \delta_c - k_c v$.

**Appendix B: Rabi frequencies $\Omega_\mu$**

The microwave magnetic field couples to the magnetic moment of the electron spin of the atom. The coupling to the nuclear magnetic moment is neglected, because it is three orders of magnitude smaller than the electron magnetic moment. The Rabi frequency of the hyperfine transition $|2, m_1\rangle \leftrightarrow |3, m_2\rangle$ us given by

(B1)
$$\Omega_{2,m_1}^{3,m_2} = \frac{2\mu_B}{\hbar} \langle 3, m_2 | B \cdot J | 2, m_1 \rangle$$

With $J = (J_{x'}, J_{y'}, J_{z'})$ the electron spin operator. Using $J_\pm = J_{x'} \pm i J_{y'}$ we can write

(B2)
$$B \cdot J = B_{x'} e^{-i\phi_{x'}} J_{x'} + B_{y'} e^{-i\phi_{y'}} J_{y'} + B_{z'} e^{-i\phi_{z'}} J_{z'}$$
$$= \frac{1}{2}\left(B_{x'} e^{-i\phi_{x'}} - i B_{y'} e^{-i\phi_{y'}}\right) J_+$$
$$+ \frac{1}{2}\left(B_{x'} e^{-i\phi_{x'}} + i B_{y'} e^{-i\phi_{y'}}\right) J_- + B_{z'} e^{-i\phi_{z'}} J_{z'}$$

Evaluating the matrix elements for the there transition connecting to $|2,0\rangle$

(B3)
$$\Omega_- = \Omega_{2,0}^{3,-1} = \frac{2\mu_B}{\hbar} \langle 3,-1 | \frac{1}{2}\left(B_{x'} e^{-i\phi_{x'}} + i B_{y'} e^{-i\phi_{y'}}\right) J_- | 2,0 \rangle$$

(B4)
$$\Omega_\pi = \Omega_{2,0}^{3,0} = \frac{2\mu_B}{\hbar} \langle 3,0 | B_{z'} e^{-i\phi_{z'}} J_{z'} | 2,0 \rangle$$

(B5)
$$\Omega_- = \Omega_{2,0}^{3,+1} = \frac{2\mu_B}{\hbar} \langle 3,1 | \frac{1}{2}\left(B_{x'} e^{-i\phi_{x'}} - i B_{y'} e^{-i\phi_{y'}}\right) J_+ | 2,0 \rangle$$